\documentclass{sf2a-conf2019}
\usepackage{graphicx}
\usepackage{hyperref}
\usepackage[]{natbib}  
\usepackage{epstopdf}
\usepackage{amsmath,amsthm,amssymb}
\usepackage{mathtools}

\def\BibTeX{{\rm B\kern-.05em{\sc i\kern-.025em b}\kern-.08em
    T\kern-.1667em\lower.7ex\hbox{E}\kern-.125emX}}
\bibpunct{(}{)}{;}{a}{}{,}  

\begin{document}

\TitreGlobal{SF2A 2019}


\title{Electron Acceleration in the Crab Nebula}

\runningtitle{Electron Acceleration in the Crab Nebula}

\author{G. Giacinti}\address{Max-Planck-Institut f\"ur Kernphysik, Postfach 103980, 69029 Heidelberg, Germany}

\author{J.\,G. Kirk$^1$}

\setcounter{page}{1}


\maketitle


\begin{abstract}
We study electron and positron acceleration at the termination shock of a striped pulsar wind. Drift motion along the shock surface keeps either electrons or positrons ---but not both, close to the equatorial plane of the pulsar, where they are accelerated by the first-order Fermi process. Their energy spectrum is a power law, and both the X-ray flux and photon index of the Crab Nebula, as measured by NuSTAR, can be reproduced for sufficiently large downstream turbulence levels. The implication that one sign of charge is preferentially accelerated in pulsar wind nebulae is potentially important for the interpretation of the positron fraction in cosmic-rays.
\end{abstract}



\section{Introduction}
\label{Sec_Introduction}

The Crab Nebula is thought to accelerate electrons and/or positrons up to at least a PeV \citep[e.g.,][]{Buehler2014}. However, the mechanisms and sites of particle acceleration remain uncertain. The photon index of the Nebula in X-rays~\citep{NuSTAR2015} is very close to expectations for electrons accelerated by the first order Fermi mechanism at an ultra-relativistic shock with isotropic particle scattering \citep{bednarzostrowski98,kirketal00}. However, the magnetic field is expected to be toroidal close to the pulsar wind termination shock (TS), i.e., the TS is perpendicular. On the one hand, diffusive shock acceleration is known to be inoperative at perpendicular shocks~\citep{begelmankirk90,sironispitkovsky09}. On the other hand, the toroidal field in the downstream region of the TS is expected to change sign across the equatorial plane of the pulsar \citep[e.g.,][]{Porth2016}: in this region, turbulence levels may be higher, and diffusive shock acceleration might still operate. We study here electron and positron acceleration in this region of the TS, by propagating individual particles in a model of the magnetic field and flow pattern \citep{GGJK2018}.

\section{Model}
\label{Sec_Model}

\begin{figure}[ht!]
 \centering
 \includegraphics[width=0.85\textwidth,clip]{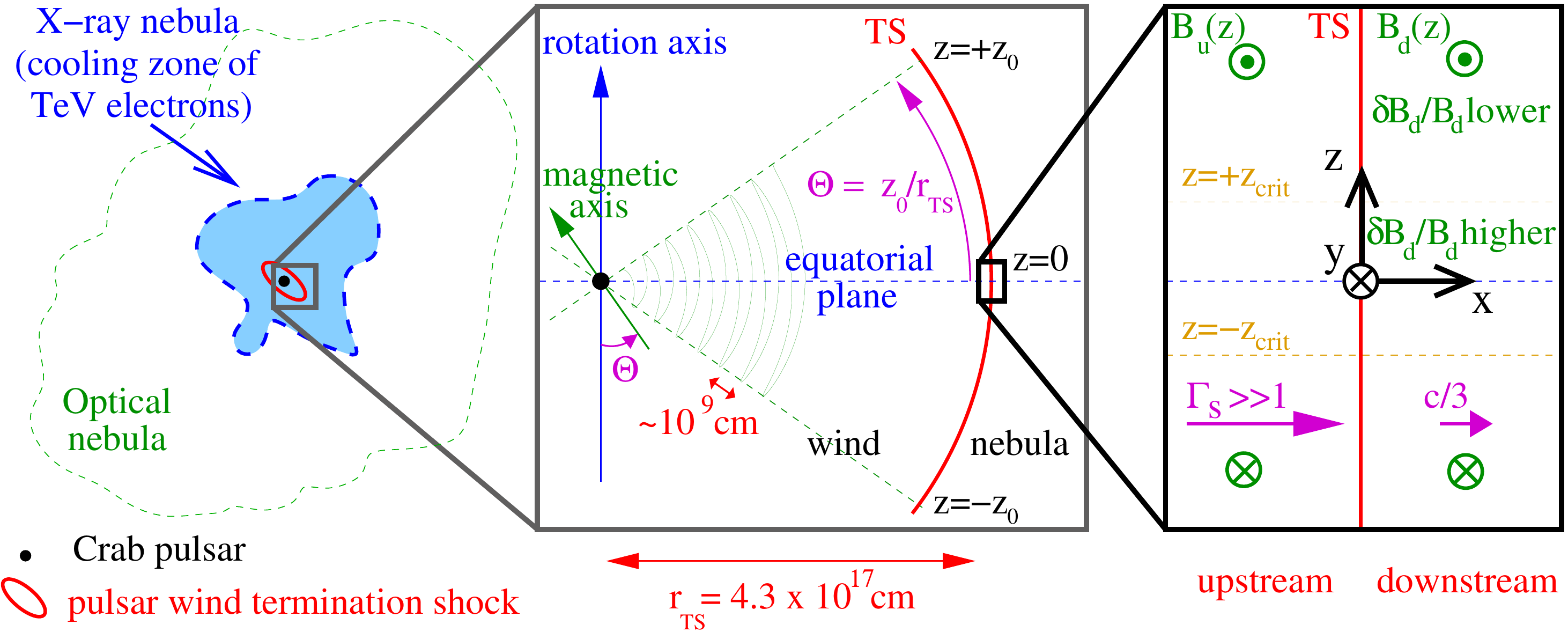}      
  \caption{Sketch of the Crab Nebula (left panel), with the location and characteristics of the region studied in this work (right panel). A description of this Figure is provided in the text in \S\ref{Sec_Model}. }
  \label{fig1}
\end{figure}

Fig.~\ref{fig1} shows a sketch of the region of interest in the Crab
Nebula. In the left panel, the X-ray and optical
nebulae are drawn as they appear on the sky, together with an estimate
of the position of the TS. The centre panel is an enlargement of the
equatorial region of the Crab pulsar wind (labelled \lq\lq
wind\rq\rq\/). The TS is drawn as a solid red arc of radius
$r_{\rm TS} \simeq 4.3 \times 10^{17}$\,cm. The rotation axis of
the pulsar (blue arrow) lies in the plane of the figure, and the
magnetic axis (green arrow) is drawn at a phase at which it lies in
this plane too. The horizontal dashed blue line corresponds to the
equatorial plane. Magnetic field oscillations, or stripes, are present
upstream of the TS between the latitudes $\pm\Theta$,
where $\Theta$ is the angle between the magnetic and rotation
axes. Upstream, the magnetic field is toroidal and 
changes sign across the current sheet (thin green line). The stripes are
destroyed at the TS, leaving a net toroidal component in the downstream
\lq\lq nebula\rq\rq\/, which reverses across the equatorial plane, and 
has an amplitude that grows with distance from this plane
\citep[e.g.,][]{Porth2016}. The black rectangle in the centre panel,
and its enlargement in the right panel correspond to the region we
model. It is typically a few percent of the shock area, so we take 
the TS to be plane and the flow planar. In the cartesian coordinate system defined in Fig.~\ref{fig1},
the fluid flows along $+{\bf \hat{x}}$. In the simulations, we set the
Lorentz factor of the fluid in the upstream ($x<0$) to $\Gamma_{\rm s}
= 100$, but the results do not depend on this choice, provided $\Gamma_{\rm s}\>>10$. 
Downstream, the fluid velocity is assumed to be
$c/3$, and the residual toroidal field, defined in the
downstream rest frame (DF), is modelled as 
${\bf B_{\rm  d}}(z) = -B_{\rm d,0} (z/z_{0}) {\bf \hat{y}}$ --- with $z_0 = \Theta r_{\rm TS}$ and $B_{\rm
  d,0} = +1$mG. The polarity of the pulsar is set by the sign of $B_{\rm d,0}$. 
Upstream, the wavelength of the stripes ($\approx 10^{9}$\,cm) is
much smaller than the gyroradii of the particles we considered, 
which, therefore, probe only the phase-averaged field, 
given by the jump conditions to be
${\bf B'_{\rm u}}(z) = (-B_{\rm
  d,0}/(2\sqrt{2}\beta_{\rm s})) \times (z/z_{0}) {\bf \hat{y}}$, 
in the shock rest frame, where $\beta_{\rm s}$ is the
velocity of the upstream fluid in this frame.  Because the downstream
flow is subsonic, turbulence is able to fill the entire Nebula
external to the TS. Therefore, on top of the residual toroidal
field, we add, in this region, a homogeneous 3D turbulent component, with
root-mean-square strength $\delta B_{\rm d}$ ---defined in the
DF. Thus, the turbulence level $\eta = \delta B_{\rm d}/|{\bf B_{\rm
    d}}|$ is larger close to the equatorial plane, in line with MHD
simulations of the Nebula \citep[e.g.,][]{Porth2016}. The turbulent
field is defined on a 3D grid with $256$ vertices per side, repeated
periodically in space, using the method of~\cite{Giacinti2012}. We use
isotropic turbulence with a Bohm spectrum, but our results do not
noticeably depend on the spectrum. For technical reasons
\citep{GGJK2018}, we add weak turbulence in the upstream, but our
results are not influenced by it. We inject electrons and positrons at
the TS with their momenta directed along $+{\bf \hat{x}}$, and with an
energy $E_{\rm inj,d}=1$\,TeV which is in the relevant range for the
Crab Nebula. We integrate the particle trajectories in the downstream
and upstream rest frames where the electric fields vanish, and do a Lorentz
transformation at each shock crossing. Defining $z_{\rm crit}$ as the
height at which the gyroradius in ${\bf B_{\rm d}}(z_{\rm crit})$ of
an injected particle is equal to $z_{\rm crit}$, one has $z_{\rm crit}
= \sqrt{z_{0}E_{\rm inj,d}/eB_{\rm d,0}} \simeq 5.8 \times
10^{14}\,{\rm cm}\sqrt{z_{0}/10^{17}\,{\rm cm}}$. The region where
injected particles are efficiently accelerated is typically $|z|
\lesssim {\rm several}~z_{\rm crit}$.

\section{Results}
\label{Sec_Results}

\begin{figure}[ht!]
 \centering
 \includegraphics[width=0.48\textwidth,clip]{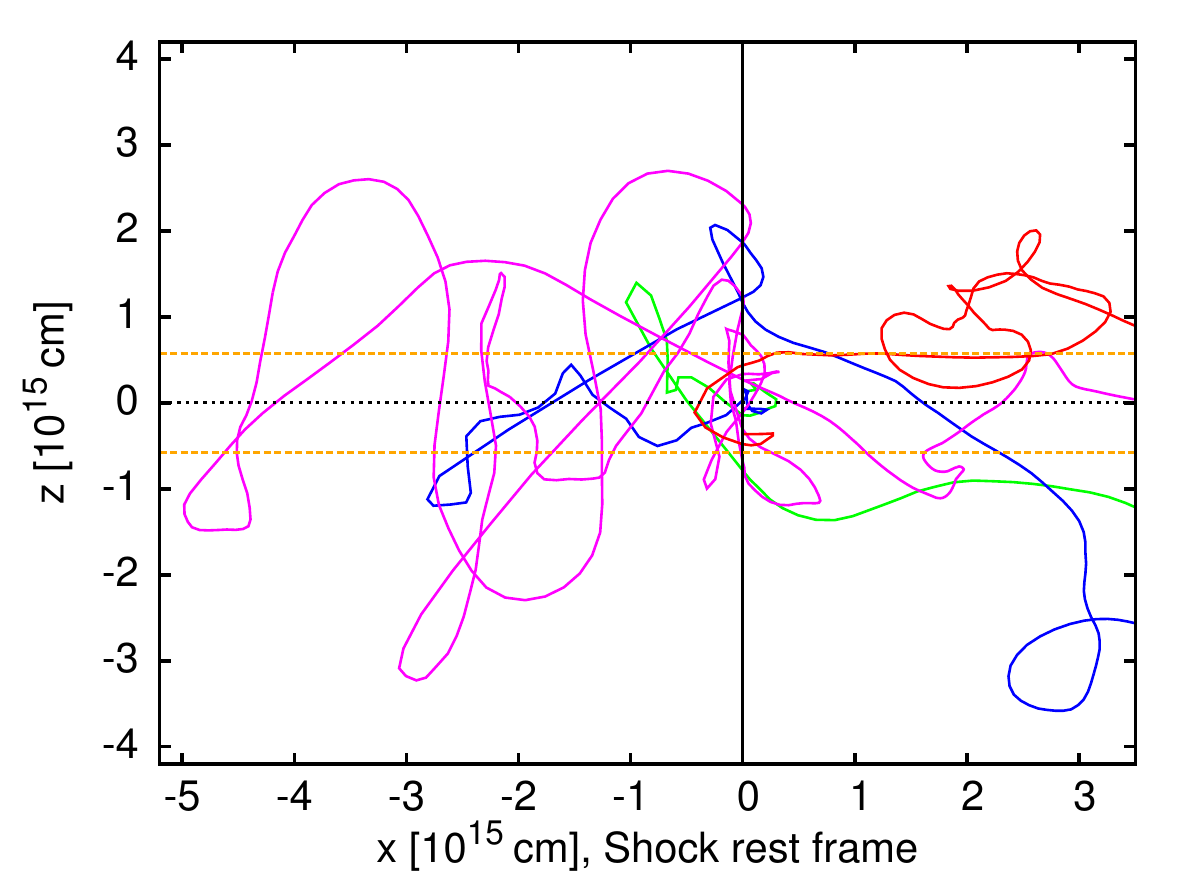}%
 \includegraphics[width=0.48\textwidth,clip]{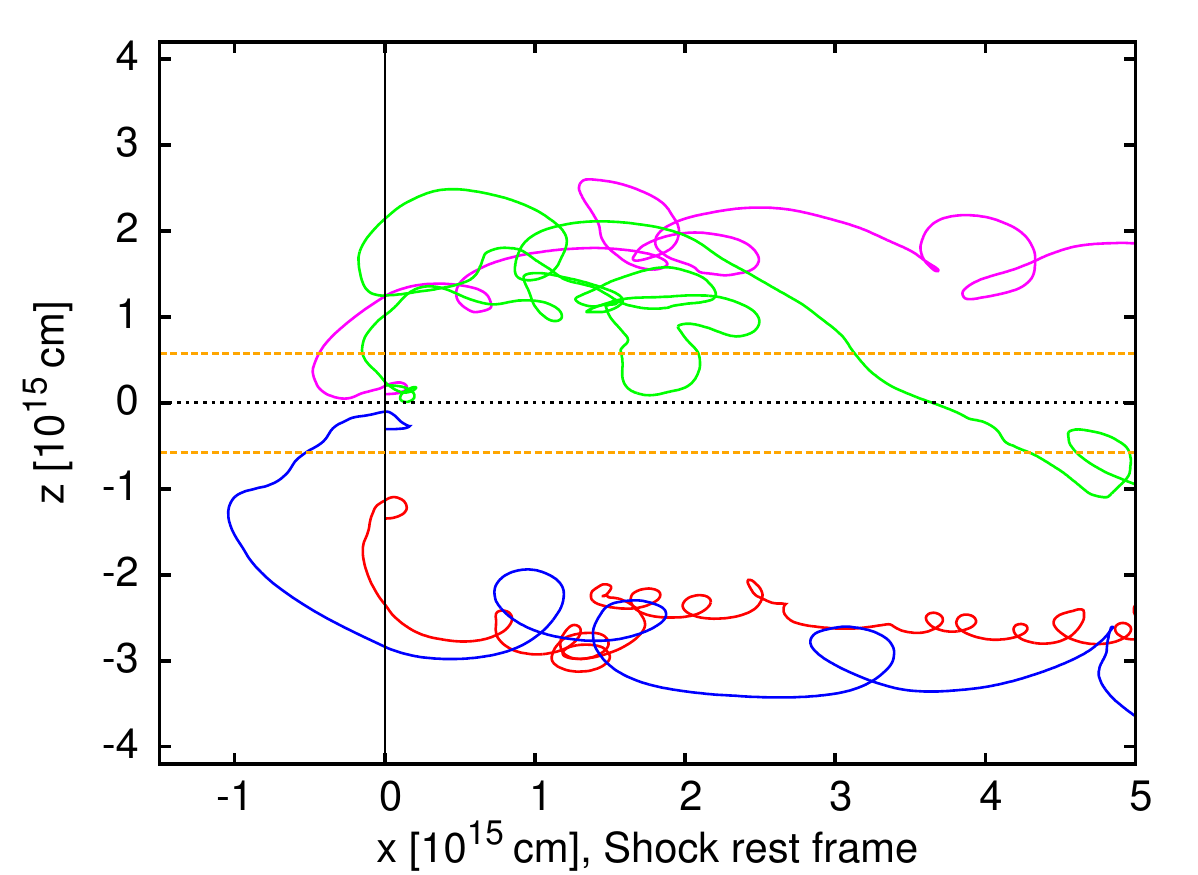}      
  \caption{{\bf Left:} Trajectories of electrons injected in the equatorial region of the TS. The upstream is on the left hand side of the shock ($x=0$, solid black line). {\bf Right:} Trajectories of positrons for the same parameters. See \S\ref{Sec_Results} for more details. }
  \label{fig2}
\end{figure}

\begin{figure}[ht!]
 \centering
 \includegraphics[width=0.48\textwidth,clip]{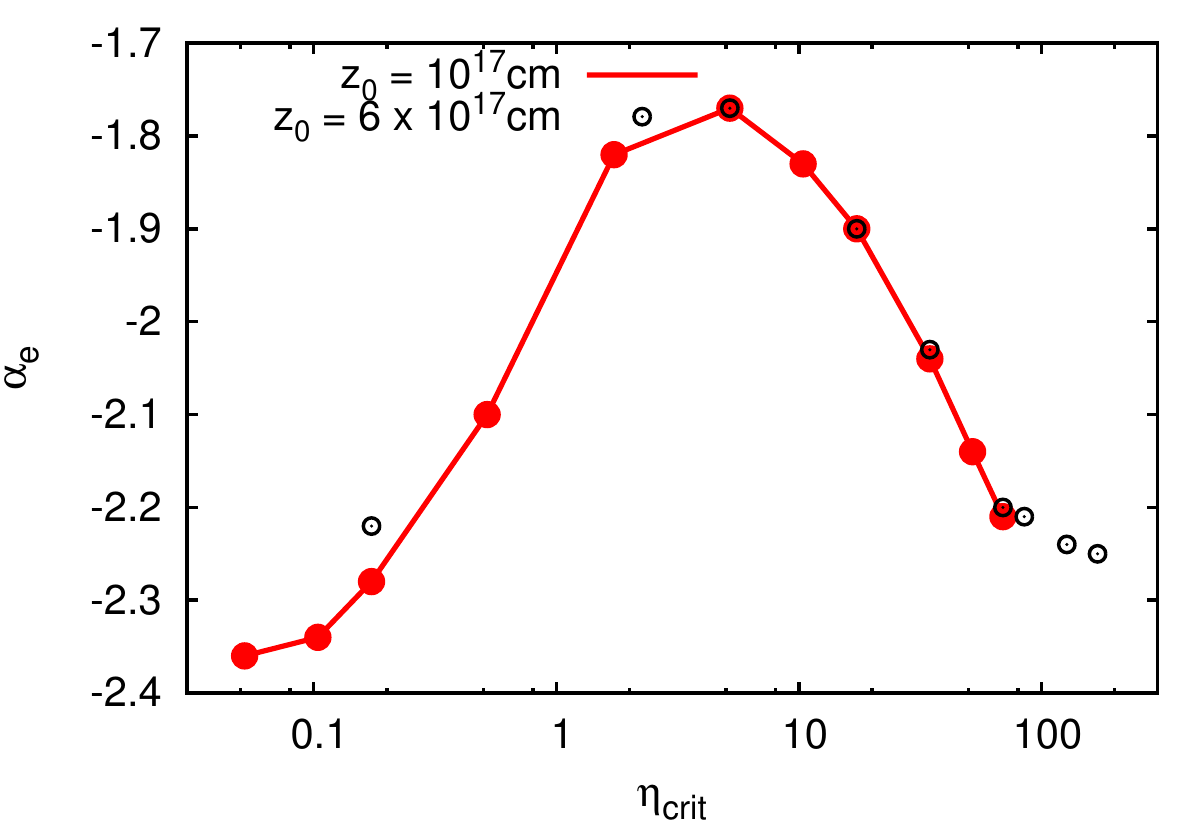}%
 \includegraphics[width=0.48\textwidth,clip]{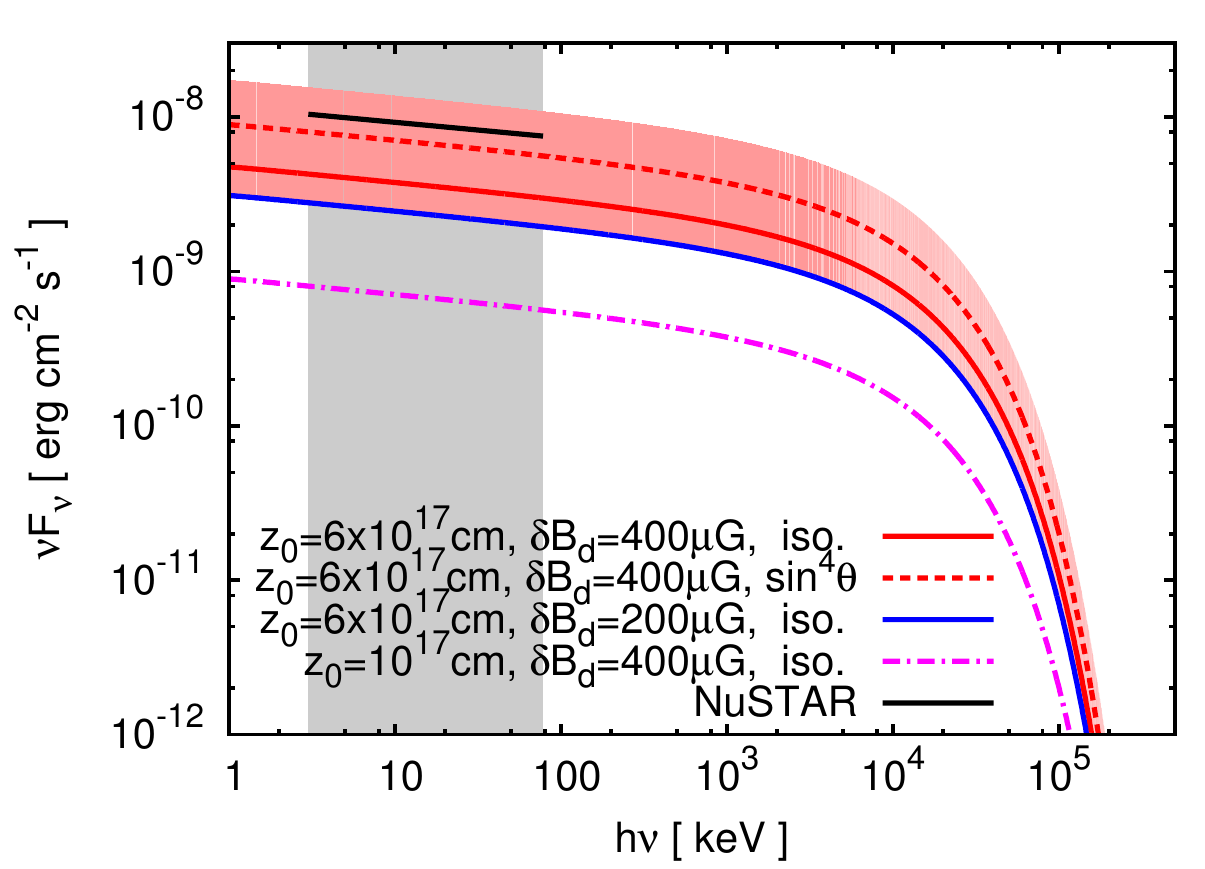}      
  \caption{{\bf Left:} Spectral index of the accelerated electrons, $\alpha_{\rm e}$, as a function of $\eta_{\rm crit}$. {\bf Right:} Predicted synchrotron spectra for the Crab Nebula versus NuSTAR measurements. See the text in \S\ref{Sec_Results} for explanations. }
  \label{fig3}
\end{figure}

We show in Fig.~\ref{fig2} the trajectories of 4 electrons (left panel) and 4 positrons (right) injected at the TS ($x=0$, solid black lines) at $|z| < 1.5\times 10^{15}$\,cm, and accelerated via the first-order Fermi process, in a simulation where $z_{0} = 10^{17}$\,cm and $\delta B_{\rm d} = 30\,\mu$G. The trajectories are plotted in the shock rest frame and projected onto the ($x$,$z$) plane. The dashed orange lines represent $z = \pm z_{\rm crit}$. In the simulations, about 90 to 95\% of injected particles are advected in the downstream without gaining energy. By comparing the two panels, one can see that the two signs of charge behave differently. For this pulsar polarity ($B_{\rm d,0}>0$), electrons are focused towards the equatorial plane (black dotted line), whereas positrons tend to be pushed away from it. This is due to drift motion on the shock surface: electrons entering the upstream at an altitude $z_1$ tend to come back in the downstream at $z_2$ such that $|z_2|<|z_1|$, whereas positrons tend to re-enter the downstream at $|z_2|>|z_1|$. Since turbulence levels are larger at small $|z|$, electrons remain confined in the region which is the most favourable for diffusive shock acceleration. A number of them stay on \lq\lq Speiser\rq\rq\/ orbits (e.g., the magenta trajectory in the left panel), and cross and re-cross the TS many times. In contrast, positrons are pushed away from this favourable region, and their acceleration quickly stops. Therefore, only electrons are accelerated to very high energies. For the opposite pulsar polarity ($B_{\rm d,0}<0$), the situation would be the opposite. The energy spectrum of the particles that are efficiently accelerated is a power-law $dN/dE \propto E^{\alpha_{\rm e}}$, where $\alpha_{\rm e}$ depends on the downstream turbulence level and lies in the range $\simeq -1.8$ to $-2.4$. We plot in Fig.~\ref{fig3} (left panel) the values of $\alpha_{\rm e}$ obtained in our simulations, versus $\eta_{\rm crit} \equiv \delta B_{\rm d}/|{\bf B_{\rm d}}|\,| _{z = z_{\rm crit}}$, for $z_0 = 10^{17}$\,cm (solid red line) and $z_0 = 6 \times 10^{17}$\,cm (black circles). We find that $\alpha_{\rm e}$ is a function of $\eta_{\rm crit}$ and does not depend on $z_0$. For low levels of turbulence $\eta_{\rm crit} \approx 1 - 30$, the spectrum is harder than $E^{-2}$. It softens to $E^{-2.2}$ for larger turbulence levels, which corresponds to the slope that is required to explain the X-ray photon index of the Crab Nebula as measured by NuSTAR. At $\eta_{\rm crit} < 1$, the spectrum also softens, but too few electrons are accelerated to explain the X-ray flux from the Nebula.

Electrons accelerated at the TS are advected in the Nebula where they cool. Assuming that the maximum electron energy at the TS is equal to 1\,PeV ---as would be expected if it is limited by synchrotron losses in a typical magnetic field strength of $\sim 0.5$\,mG, we calculate the synchrotron spectrum from the cooled electrons and plot the results in Fig.~\ref{fig3} (right panel), for four sets of parameter values. See the key for the values of $z_0$ and $\delta B_{\rm d}$, and for the isotropy (\lq\lq iso.\rq\rq\/) or anisotropy of the pulsar wind. We consider both isotropic and $\propto \sin^{4}\theta$ winds, where $\theta$ denotes the colatitude. We use 2.0\,kpc for the distance to the Crab pulsar. The effect of the uncertainty on this distance ($\pm 0.5$\,kpc) for the two red lines is shown with the area shaded in red. The solid black line corresponds to the measurements from NuSTAR in the $3-78$\,keV band \citep{NuSTAR2015}. Our model can reproduce them for sufficiently large values of $\delta B_{\rm d}$ ($\geq 200\,\mu$G) and $z_0$. The magenta dashed-dotted line for $z_0 = 10^{17}$\,cm (i.e., $\Theta \simeq 13^{\circ}$) and $\delta B_{\rm d} = 400\,\mu$G is about an order magnitude below the measurements, but we obtain a larger X-ray flux for $z_0 = 6 \times 10^{17}$\,cm (i.e., $\Theta \simeq 80^{\circ}$): the blue and red solid lines correspond to $\delta B_{\rm d} = 200\,\mu$G and $\delta B_{\rm d} = 400\,\mu$G for an isotropic wind. The red dashed line is for a $\propto \sin^{4}\theta$ wind and $\delta B_{\rm d} = 400\,\mu$G. We can reproduce the data with these parameters. $|{\bf B}_{\rm d}|(z) \propto |z|$ here, and the measurements would be reproduced with smaller values of $\Theta$ and $\delta B_{\rm d}$, if one adopts a shallower dependence of $|{\bf B}_{\rm d}|$ on $z$.

\section{Discussion}
\label{Sec_Discussion}

We find that the acceleration of X-ray emitting electrons occurs preferentially in the equatorial region of the TS. Interestingly, modeling of the high-energy emission from the Crab Nebula is compatible with these electrons being accelerated in, or close to, this region \citep{Olmi2016}. Shock-drift plays an important role, and ensures that the accelerated electrons remain in the equatorial region of the TS. For sufficiently large turbulence levels, the electron spectral index tends towards $-2.2$, which is compatible with theoretical expectations \citep[e.g.,][]{bednarzostrowski98}. For lower turbulence levels, the spectral index increases up to $-1.8$. This may explain the hard photon index measured in the central regions of the Nebula by the Chandra X-ray Observatory \citep{Mori2004}, as turbulence levels may vary with time and position at the TS. We note that other effects, such as shock corrugation \citep{Lemoine_JPlPh2016}, may also play a role in the acceleration of X-ray emitting electrons, and that another acceleration mechanism may operate upon the electrons responsible for the radio to optical emission of the Nebula \citep{Olmi2016,LyutikovSironi2019}. The gamma-ray flares detected by AGILE and Fermi-LAT from the Crab Nebula require another acceleration mechanism too, such as inductive acceleration in the striped wind~\citep{Kirk2017,Kirk2019}. Finally, the fact that each pulsar may accelerate preferentially either electrons or positrons to high energy, but not both, could have important implications for the interpretation of the positron fraction in cosmic-rays. Studies usually assume that pulsars accelerate electrons and positrons in equal numbers. Under this assumption, the fact that the AMS-02 positron fraction saturates well below 0.5 seems to rule out nearby pulsars as the main source of the high-energy electrons and positrons detected at Earth \citep{Recchia_AMS02}. However, our above findings show that pulsars do remain viable candidates, as long as the local pulsar(s) responsible for these fluxes accelerate preferentially electrons rather than positrons.

\section{Conclusions}
\label{Sec_Conclusions}

We study particle acceleration at the TS of a striped pulsar wind. We find that either electrons or positrons are accelerated to very high energy, depending on the relative orientations of the magnetic and rotation axes of the pulsar. Drift motion on the shock surface keeps the accelerated particles close to the equatorial plane of the pulsar, allowing them to be accelerated by the first order Fermi process at the TS. Their energy spectrum is a power law, with index in the range $-1.8$ to $-2.4$. Both the X-ray flux and photon index of the Crab Nebula, as measured by NuSTAR, can be reproduced for sufficiently large turbulence levels downstream of the shock. Our results strongly question the assumption often used in studies of the positron fraction that pulsars accelerate electrons and positrons to high energy in equal numbers.

\begin{acknowledgements}
This research was supported by a Grant from the GIF, the German-Israeli Foundation for Scientific Research and Development.
\end{acknowledgements}

\bibliographystyle{aa}  
\bibliography{giacinti_S08} 

\end{document}